\documentclass[aps,prl,10pt,twocolumn,showpacs,superscriptaddress]{revtex4-1}
\usepackage{graphicx}
\usepackage{amsmath}
\usepackage{amssymb}
\usepackage{psfrag}
\usepackage{natbib}
\usepackage{color,xcolor}

\def \w {\omega}
\def \W {\Omega}

\begin{document}
\title{Chimera patterns in the Kuramoto-Battogtokh model}
\author{L. A. Smirnov}
\affiliation{Department of Control Theory, Nizhny Novgorod State University,
Gagarin Av. 23, 606950, Nizhny Novgorod, Russia}
\affiliation{Institute of Applied Physics, Ul'yanov str.\,46, 603950, Nizhny Novgorod, Russia}
\author{G. V. Osipov}
\affiliation{Department of Control Theory, Nizhny Novgorod State University,
Gagarin Av. 23, 606950, Nizhny Novgorod, Russia}
\author{A. Pikovsky}
\affiliation{Institute for Physics and Astronomy, 
University of Potsdam, Karl-Liebknecht-Str. 24/25, 14476 Potsdam-Golm, Germany}
\affiliation{Department of Control Theory, Nizhny Novgorod State University,
Gagarin Av. 23, 606950, Nizhny Novgorod, Russia}
\begin{abstract}
Kuramoto and Battogtokh [Nonlinear Phenom. Complex Syst. 5, 380 (2002)] discovered 
chimera states represented by stable coexisting synchrony and asynchrony domains in a
lattice of coupled oscillators. After reformulation in terms of local order parameter, the problem
can be reduced to partial differential equations. We find uniformly rotating periodic in space
chimera patterns as solutions of a reversible ordinary differential equation, and demonstrate 
a plethora of such states. In the limit of neutral coupling they reduce to
analytical solutions in form of one- and two-point chimera patterns as well as 
localized chimera solitons. Patterns at weakly attracting coupling are 
characterized by virtue of a perturbative approach. Stability analysis reveals that only simplest chimeras 
with one synchronous region are stable.
\end{abstract}
\maketitle
\looseness=-1
Chimera states in populations of coupled oscillators have attracted large
interest since their first observation and theoretical explanation
by Kuramoto and Battogtokh~\cite{Kuramoto-Battogtokh-02}. The essence 
of chimera is in the spontaneous symmetry breaking: although a homogeneous
fully symmetric synchronous state exists, yet another nontrivial state combining
synchrony and asynchrony  is  possible and can even be stable. Chimeras
can be found at interaction of several populations of 
oscillators~\cite{Abrams-Mirollo-Strogatz-Wiley-08,*Pikovsky-Rosenblum-08,%
*Tinsley_etal-12,*Martens_etal-13}, or
in an oscillatory 
medium~\cite{Abrams-Strogatz-04,ShiKu04,Laing-09,Bordyugov-Pikovsky-Rosenblum-10}, 
the latter situation can be treated as a pattern formation
problem. Here the formulation in terms of the
coarse-grained complex order parameter indeed
allows one to reduce the problem to that of evolution of a complex 
field~\cite{Laing-09,Bordyugov-Pikovsky-Rosenblum-10}. For a recent review see~\cite{Panaggio-Abrams-15}.

The goal of this paper is to develop a theory of chimera patterns in a one-dimensional (1D)
medium based on formulation of the problem as a set of partial differential equations (PDEs). This
allows us to represent
the chimera state as a solution of ordinary differential equations (ODEs), periodic in space
chimeras correspond to periodic orbits of these ODEs.
We show that in a limit of neutral coupling, these equations are integrable yielding  singular
``one-point'' and ``two-point''  chimeras; for a weakly attracting coupling we find properties of chimera patterns
by virtue of perturbation analysis to these solutions. 
Furthermore, we study stability of found chimera patterns
by employing a numerical method allowing to disentangle the essential continuous and 
the discrete (point)
parts~\cite{Wolfrum_etal-11,Omelchenko-13} of the stability spectrum.\par
The original Kuramoto-Battogtokh (KB) model~\cite{Kuramoto-Battogtokh-02} is formulated as a 1D field of phase oscillators $\phi\left(x,t\right)$ evolving according to
\begin{equation}
\partial_{t}\phi=
\w+\mathrm{Im}\left(e^{-i\left(\phi+\alpha\right)}\!\int\!G\left(x-\tilde{x}\right) e^{i\phi(\tilde{x},t)}d\tilde{x}\right),
\label{eq:kbmodel}
\end{equation}
with exponential kernel  $G\!\left(y\right)=\kappa \exp\bigl(-\kappa\left|y\right|\bigr)\!\bigl/2\bigr.$. Coupling 
is attractive if the phase shift $\alpha<\pi\bigl/2\bigr.$, then the synchronous state
where all the phases are equal is stable; $\alpha=\pi\bigl/2\bigr.$ corresponds to neutral coupling.  

One can reformulate this setup as a 1D continuous oscillatory 
medium~\cite{Laing-09,Bordyugov-Pikovsky-Rosenblum-10}, 
described  by the complex field $Z\!\left(x,t\right)$, which represents a coarse-grained order parameter of the phases:
$Z\!\left(x,t\right)=\frac{1}{2\delta}\int_{x-\delta}^{x+\delta} e^{i\phi(\tilde{x},t)}d\tilde{x}$.
In the synchronous state $|Z|=1$, while for partial synchrony $0<|Z|<1$.
The  dynamics of $Z\!\left(x,t\right)$ just follows locally the Ott-Antonsen equation~\cite{Ott-Antonsen-08,Panaggio-Abrams-15}
\begin{equation}
\partial_{t}Z=i\w Z+\bigl(e^{-i\alpha}H-e^{i\alpha}H^{\ast}Z^{2}\bigr)\bigl/2\bigr.\;.
\label{eq:oa1}
\end{equation}
where a field $H\!\left(x,t\right)\!=\!\int \!G\!\left(x-\tilde{x}\right)Z\!\left(\tilde{x},t\right)d\tilde{x}$ describes~the force due to coupling.
This nonlocal coupling according stems from the following model 
for the interaction of oscillators via the ``auxiliary''
field $H$ (cf. Refs.~\cite{ShiKu04,Laing-11,Laing-15}):
\begin{equation}
\tau\partial_{t}H=\kappa^{-2}\partial^{2}_{xx}H-H+Z\;.
\label{eq:difeq1}
\end{equation}
In the limit $\tau\to0$, Eq.~\eqref{eq:difeq1} reduces to an equation
\begin{equation}
\partial^{2}_{xx}H-\kappa^2 H =-\kappa^2 Z\;,
\label{eq:difeq2}
\end{equation}
solution of which depends on boundary conditions.  In particular, in an infinite medium $\left|x\right|<\infty$, 
the solution is $H\!\left(x,t\right)=\!\int \!\left(\kappa \exp\bigl(-\kappa\left|x-\tilde x\right|\bigr)\!\bigl/2\bigr.\right)
Z\!\left(\tilde{x},t\right)d\tilde{x}$ as in~\eqref{eq:kbmodel}.

Below we consider periodic in space medium with period $L$, in this case 
the KB model exactly corresponds to Eqs.~\eqref{eq:oa1},{\,}\eqref{eq:difeq2} if integration 
is performed in an infinite domain while the fields are assumed to have period $L$. 
If an integration over the periodic domain of size $L$ is performed,
one should use the kernel
\begin{equation}
G(y)=\frac{\kappa}{2\sinh\bigl(\kappa L\bigl/2\bigr.\bigr) }\cosh\bigl(\kappa\left(\left|y\right|-L\bigl/2\bigr.\right)\bigr)\;,
\label{eq:kernel}
\end{equation}
which follows from the solution of  \eqref{eq:difeq2} with periodic boundary conditions.

The formulated problem \eqref{eq:oa1},{\,}\eqref{eq:difeq2} contains two parameters having dimension of 
length: $\kappa$ and $L$.
By rescaling coordinate $x$ we can set one of these parameters to one. 
It is convenient to set $\kappa=1$, then the only parameter
is the size of the system $L$.\par
Our next goal is to find chimera states, which consist of synchronous and asynchronous parts. We look for rotating-wave solutions of system \eqref{eq:oa1},{\,}\eqref{eq:difeq2}, which are stationary
in a rotating reference frame:
$Z(x,t)=z(x)e^{i(\w+\W)t}$, $H(x,t)=h(x)e^{i(\w+\W)t}$,
where $\W$ is some unknown frequency to be defined below~\cite{bib:cpkbm_note}. Substituting this we get a system of an algebraic equation and  an ODE for complex functions $z\!\left(x\right)$ and $h\!\left(x\right)$
\begin{gather}
e^{i\alpha}h^{\ast}z^2+2i\W z-e^{-i\alpha}h=0\;,\label{eq:zeq}\\
h''-h=-z\;.\label{eq:heq}
\end{gather}
 Here and below prime denotes spatial derivative.

The first step is to express $z\!\left(x\right)$ from the quadratic equation~\eqref{eq:zeq}~(see~\cite{bib:sm}). This equation describes the order parameter $z\!\left(x\right)$ of
a set of oscillators driven by field $h\!\left(x\right)=r\!\left(x\right)e^{i\theta\left(x\right)}$, the solution at each point $x$ depends on the relation between $r$ and $\W$ (below for
simplicity of presentation we write the relations for $\W<0$). If $\left|r\right|\geq\left|\W\right|$, then the oscillators are locked and $|z|=1$, otherwise
the oscillators are partially synchronous with $0<|z|<1$. The solution reads
\begin{equation}
z\!=\!\begin{cases} -\left(i\W-\sqrt{r^2\!-\!\W^2}\right)r^{-1}e^{i(\theta-\alpha)},\hspace{2mm}\text{if}\hspace{2mm} \left|r\right|\!\geq\!\left|\W\right|,\vspace{1.0mm}\\
-i\left(\W+\sqrt{\W^2\!-\!r^2}\right)r^{-1}e^{i(\theta-\alpha)},\hspace{2mm}\text{if}\hspace{2mm} \left|r\right|\!<\!\left|\W\right|.
\end{cases}\label{eq:sas}
\end{equation}
We now substitute this solution in~\eqref{eq:heq}. Although $h\!\left(x\right)$ is complex, the resulting equation can be written,
due to gauge invariance $\theta\!\left(x\right)\!\to\!\theta\!\left(x\right)+\theta_{0}$,  as a third-order system of ODEs for real functions $r\left(x\right)$ and $q\left(x\right)\!=\!r^2\left(x\right)\theta'\left(x\right)$
\begin{equation}
\begin{aligned}
r''&=r+r^{-3}q^{2}\!-r^{-1}\sqrt{r^2-\W^2}\cos\alpha+r^{-1}\W\sin\alpha\;,\\
q'&=\W\cos\alpha+\sqrt{r^2-\W^2}\sin\alpha\;,
\end{aligned}
\label{eq:rqsyn}
\end{equation}
in the domain where $\left|r\right|\!\geq\!\left|\W\right|$, and
\begin{equation}
\begin{aligned}
r''&=r+r^{-3}q^2+r^{-1}(\W+\sqrt{\W^2-r^2})\sin\alpha\;,\\
q'&=(\W+\sqrt{\W^2-r^2})\cos\alpha\;,
\end{aligned}
\label{eq:rqasyn}
\end{equation}
in the domain where $\left|r\right|\!<\!\left|\W\right|$.

Our goal is to find chimera patterns described by equations~\eqref{eq:rqsyn},{\,}\eqref{eq:rqasyn}, satisfying periodicity
condition $r(x+L)=r(x)$, $q(x+L)=q(x)$. It is more convenient not to fix the period $L$, but to fix the frequency of the rotating
chimera $\W$, then to find periodic solutions of \eqref{eq:rqsyn},{\,}\eqref{eq:rqasyn}, period $L$ of which depends on $\W$ (see~\cite{bib:sm}). This will
after inversion yield dependence $\W(L)$.

\begin{figure}[tb]
\centering
\includegraphics[width=\columnwidth]{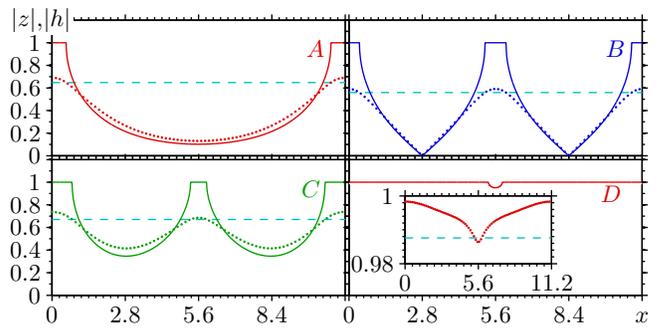}
\caption{(Color online) Profiles of simplest chimeras  (with at most two SRs) for $\alpha\!=\!1.457$, $|z|$: solid lines, $|h|$: dotted lines.
$A$: the KB one-SR chimera for $\W\!=\!-0.648$,
$B$: symmetric two-SRs chimera for $\W\!=\!-0.558$,
$C$: asymmetric two-SRs chimera (here the widths of synchronous domains are different, and their phases differ not by $\pi$, like in case $B$) for $\W\!=\!-0.672$,
$D$: nearly synchronous one-SR chimera for $\W\!=\!-0.98762$.
Colors used correspond  to coding of solutions in Fig.~\ref{fig:per}.}
\label{fig:ill}
\end{figure}

\begin{figure}[tb]
\centering
\includegraphics[width=0.7\columnwidth]{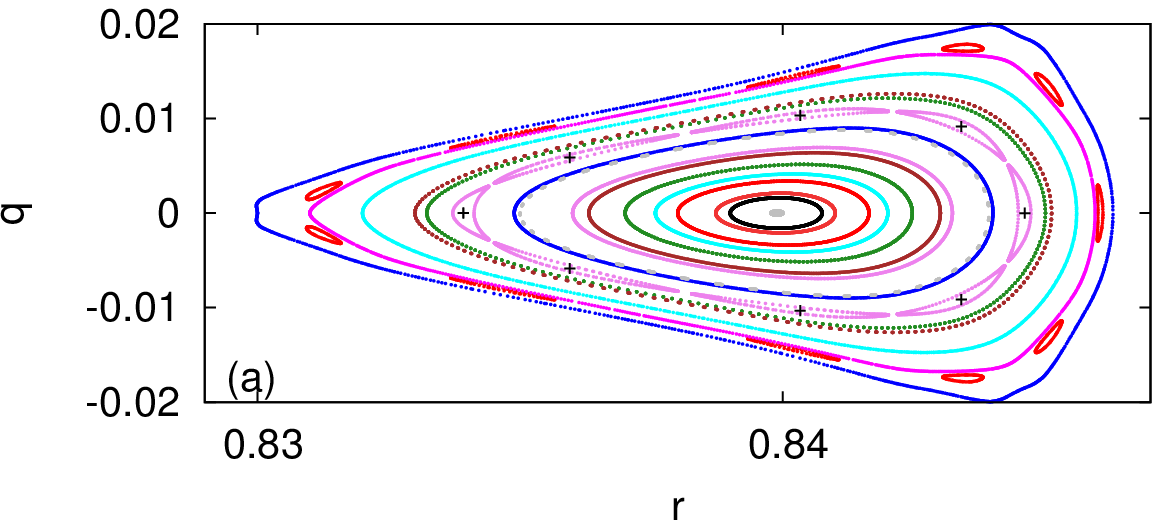}
\includegraphics[width=0.7\columnwidth]{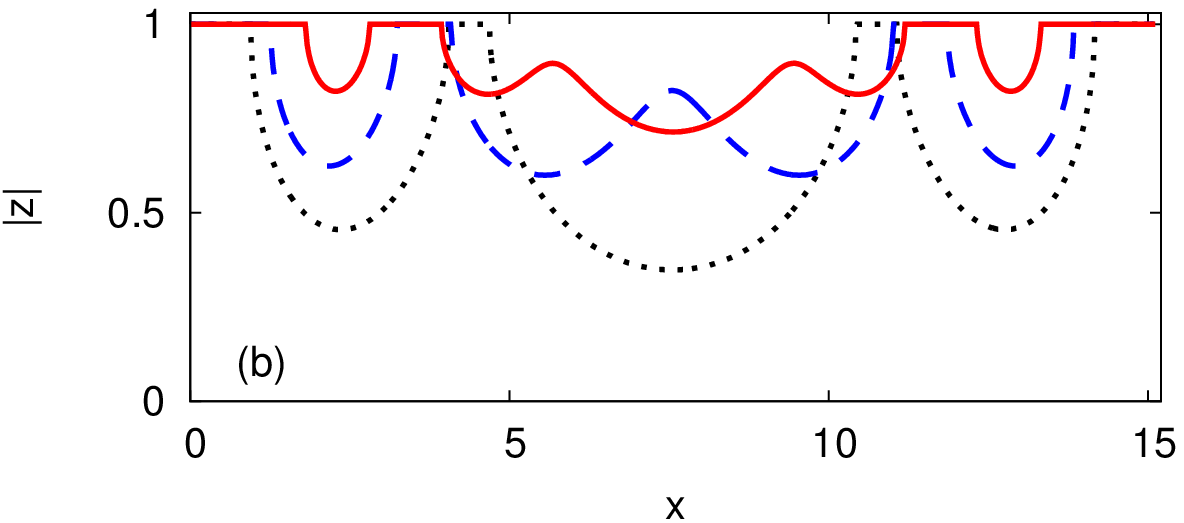}
\caption{(Color online) (a) Poincar\'e map for system  \eqref{eq:rqsyn},{\,}\eqref{eq:rqasyn} 
for $\alpha=1.457$ and $\W=-0.8$. 
The condition for the section:
$r'=0$, $r''<0$. (b) More complex patterns with three SRs
for $L\approx 15.1$ and $\W=-0.796$ (solid red line), $\W=-0.726$ (dashed blue line) and $\W=-0.674$ (dotted black line).}
\label{fig:pm}
\end{figure}

Before discussing numerical and analytical approaches, we illustrate in 
Fig.~\ref{fig:ill} several solutions
for $\alpha=1.457$ (the value used in~\cite{Kuramoto-Battogtokh-02}) with 
period $L\approx 11.2$. Presented solutions (types $A$ and $B$ have been already discussed in the 
literature~\cite{Kuramoto-Battogtokh-02,Omelchenko-13,Panaggio-Abrams-15})
are just simplest possible chimeras
with at most two synchronous regions (SRs). Indeed, 
the system \eqref{eq:rqsyn},{\,}\eqref{eq:rqasyn} is
a reversible (with respect to involution $r\to r$, $q\to -q$)  third-order system of ODEs with a plethora of 
solutions, including
chaotic ones. We illustrate this by constructing a two-dimensional Poincar\'e map in Fig.~\ref{fig:pm}a. It 
shows typical for 
nearly integrable Hamiltonian
systems picture of tori and periodic orbits of different periods. Not all points on the Poincar\'e surface lead to
physically meaningful solutions: we discarded the trajectories which resulted in values  $\left|r\right|\!>\! 1$.  The fixed point of the map Fig.~\ref{fig:pm}(a)
at $q=0$, $r\approx 0.84$ describes the one-hump chimera state $A$ in Fig.~\ref{fig:ill}. 
The Poincar\'e map is constructed for a fixed value of $\W$, it provides several branches
of periodic orbits having different periods. Collecting solutions at a fixed period $L$, we obtain are many coexisting 
chimera patterns; several three-SRs chimeras
are illustrated in Fig.~\ref{fig:pm}b. Our aim 
in this study is not to follow
all possible periodic and chaotic solutions of this reversible system, below we focus on the simplest ones
illustrated in Fig.~\ref{fig:ill}, 
corresponding to fixed
points and period-two orbits of the Poincar\'e map.

\begin{figure}[tb]
\centering
\includegraphics[width=\columnwidth]{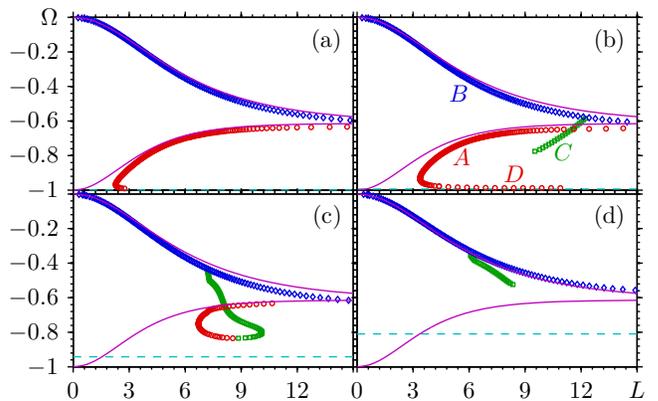}
\caption{(Color online) Periods of chimera states $L$ vs. parameter $\W$ for $\alpha=1.514$ (a), 
$\alpha=1.457$ (b), $\alpha=1.229$ (c), and $\alpha=0.944$ (d). 
Chimera states for $\alpha=\pi\bigl/2\bigr.$, obtained by integration 
Eq.~\eqref{eq:pot}, are shown with violet solid lines. Different
markers correspond to chimera types depicted in Fig.~\ref{fig:ill}, as specified in panel (b). Cyan dashed lines 
show frequency of the synchronous state $\W=-\sin\alpha$.}
\label{fig:per}
\end{figure}

Remarkably, it is possible to describe basic chimera profiles semi-analytically, for 
$\alpha\approx \pi\bigl/2\bigr.$.
Let us first consider the limiting case $\alpha=\pi\bigl/2\bigr.$. Here,
according to~\eqref{eq:rqsyn},{\,}\eqref{eq:rqasyn}, the derivative $q'\!\left(x\right)$
is non-negative in the synchronous state   and vanishes
in the asynchronous state. Thus, a periodic solution with 
$q(x)=q(x+L)$ should be everywhere asynchronous, except possibly for one
or two points at which $r(x)$ achieves an extremum $|r|=|\W|$. For this degenerate
chimera Eqs.~\eqref{eq:rqasyn} reduce to $q=0$ and an integrable second-order equation
\begin{equation}
\begin{gathered}
r''=-dU(r)\bigl/dr\bigr.\;,\\
U(r)=-r^2\bigl/2\bigr.-\sqrt{\W^2\!-\!r^2}-\W\ln\!\left(\sqrt{\W^{2}\!-\!r^{2}}\!-\!\W\right).
\end{gathered}
\label{eq:pot}
\end{equation}
In the potential $U\!\left(r\right)$, there are two types of trajectories, having maximum at $r_{max}=|\W|$, depending on the value of $\W$.
For $-1<\W<\W_{\ast}=2\left(\ln2-1\right)$ this is a periodic orbit with $0<r_{min}\leq |\W|$.
It reaches the boundary of the asynchronous region at one point and corresponds
to ``one-point chimera'',  which can be considered as the limiting case
of curve $A$ in Fig.~\ref{fig:ill}, where the SR shrinks to a point.
For $\W_{\ast}<\W<0$ there is a symmetric periodic orbit (here it is convenient
to allow $r$ to change sign, this corresponds to a jump by $\pi$ in phase $\theta$
if $r$ is considered as positive like in Fig.~\ref{fig:ill}, curve $B$) with
$-|\W|\leq r \leq |\W|$. This ``two-point chimera'' corresponds to curve $B$ in
Fig.~\ref{fig:ill}. These two types of solutions merge in a homoclinic orbit with
infinite period at  $\W=\W_{\ast}$, which can be named ``chimera soliton'' (one- or two-point,
depending on which side of the threshold the orbit is considered). The dependencies $\W\!\left(L\right)$ 
for these solutions are shown in Fig.~\ref{fig:per} as solid lines.
Note that, additionally, there is 
a branch of synchronous solutions with $\W=-1$ which are steady states $r=1$. 

The solutions above are degenerate chimeras, as the SR is 
restricted to one or two points. Synchronous region becomes finite for $\alpha\lesssim \pi\bigl/2\bigr.$,
here one can develop a perturbation approach by introducing a
small parameter $\beta=\pi\bigl/2\bigr.-\alpha\ll 1$. Now $q\neq 0$, but because $q\sim \beta$,
we can neglect terms $\sim q^2$ in  \eqref{eq:rqsyn},{\,}\eqref{eq:rqasyn}.
Then, the problem reduces to finding a periodic trajectory $r(x)$ of integrable equation, such that evolution of $q(x)$ is periodic:
\begin{equation}
\begin{gathered}
\hspace{-6.0mm}
\oint\!q'\!\left(x\right)dx \!=\!\!\int_{x:\,\left|r\left(x\right)\right|\geq\left|\W\right|}\! \left(\W\beta\!+\!\sqrt{r(x)^2\!-\!\W^2}\right)dx+{}\\{}\hspace{14.5mm}
\int_{x:\,\left|r\left(x\right)\right|\geq\left|\W\right|}\! \beta\!\left(\W\!+\!\sqrt{\W^{2}\!-\!r^{2}\!\left(x\right)}\right)dx\!=\!0\;.
\end{gathered}
\label{eq:intq}
\end{equation}
Detailed calculations are presented in \cite{bib:sm}. The result is that the size $L_{syn}$ of SR 
becomes finite:
\begin{equation}
L_{syn}\approx\sqrt{\frac{8\beta}{\pi N_{SR}\sqrt{|\W|(1-|\W|)}}\oint\!\left(R'^{2}+R^{2}\right)dx}\;.
\label{eq:chreg}
\end{equation}
where $R(x)$ is the solution of \eqref{eq:pot} at $\beta=0$ and $N_{SR}$ is the number of SRs.

We compare the analytical approach above with the results
of direct numerical calculation, in the framework of \eqref{eq:rqsyn},{\,}\eqref{eq:rqasyn}, 
of periodic orbits in Fig.~\ref{fig:per}, 
for several values of $\alpha$. Panel (a) shows that for small $\beta$ chimera states (of types $A$ and $B$
of Fig.~\ref{fig:ill}) are close to
degenerate regimes at $\beta=0$. One can see in panels (a,b) that the two analytic 
solutions at $\alpha=\pi\bigl/2\bigr.$ (the one-point chimera and the
synchronous state) merge into one branch at $\alpha\lesssim \pi\bigl/2\bigr.$ with a nonmonotonous
dependence $\W$ on $L$, cf. one-SR chimeras $A$ and $D$ in Fig.~\ref{fig:ill}.  In panel (b) one can see
an additional branch corresponding to the two-SRs asymmetric chimera $C$ in  Fig.~\ref{fig:ill}. 
As a result, in (b) and (c) one has four  solutions in some range of periods $L$.
Only two of them survive for  small $\alpha$; 
diagrams for $\alpha<0.9$ are qualitatively the same as panel (d) in Fig.~\ref{fig:per}.
 
\looseness=-1
Next, we discussed stability of the obtained chimera patterns. 
For this goal we linearize Eq.~\eqref{eq:oa1},{\,}\eqref{eq:kernel} (see~\cite{bib:sm}).
Contrary to the problem of finding chimera solutions, this analysis cannot be reduced
to that of differential equations, rather we have to consider an integral-differential 
equation~\eqref{eq:oa1},{\,}\eqref{eq:kernel} for 
$Z(x,t)$. After spatial discretization we get a matrix eigenvalue problem. The difficulty here
is that, according to~\cite{Omelchenko-13,Xie_etal-14}, there is an essential continuous 
$T$-shaped spectrum $\lambda_c$ consisting of 
eigenvalues on the imaginary and the negative real axis, but stability is determined by the point spectrum
$\lambda_{p}$.
Unfortunately, it is not easy to discriminate these parts of the spectrum in the eigenvalues $\lambda$
of the approximate matrix, because the eigenvalues representing essential part of spectrum lie
not exactly on the imaginary axis. 
We adopted the following procedure to select the point spectrum $\lambda_{p}$. For a chimera state in the domain
$0\leq x\leq L$, we can discretize the linearized system by using a set of points
$x_0+j \Delta$, \mbox{$j=0,1,\ldots,M-1$}, where $\Delta=L/M$ and $0\leq x_0\leq \Delta$
is an arbitrary continuous parameter.
This leads to an  $2M\times 2M$ real matrix, eigenvalues $\lambda$
of which we obtained numerically.
Additionally, we varied the offset of discretization $x_0$. In numerics we used $M=2048$
and $N=64$ or $N=128$ equidistant values of $x_0$. 
We have found that while the components of the essential
spectrum vary with $x_0$, the point spectrum $\lambda_{p}$ components vary extremely weakly with $x_0$ -- this allowed us to determine
point spectrum $\lambda_{p}$ reliably for most values of parameters. 
  
\begin{figure}[tb]
\centering
\includegraphics[width=\columnwidth]{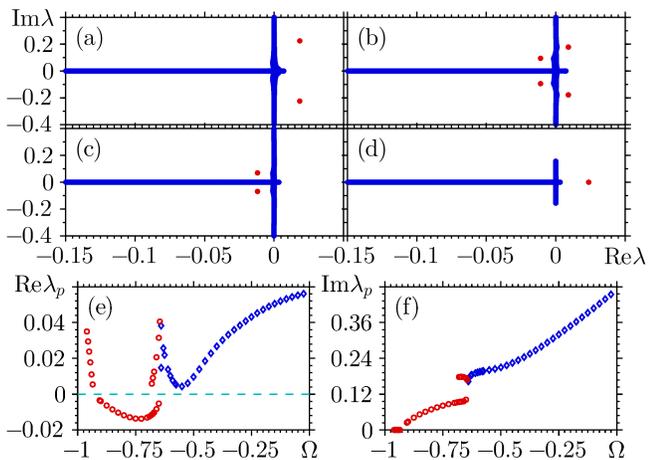}
\caption{(Color online) (a-d): Essential (blue markers) and point (red markers) spectra for chimera states at $\alpha=1.457$ and
four values of $\W$: (a) $\W=0.45$, (b) $\W=0.675$, (c) $\W=0.8$, (d) $\W=0.95$.  In these plots all $2MN$ eigenvalues
with $M=2048$ and $N=128$ are plotted. (e,f): real and imaginary parts of point spectrum $\lambda_{p}$ for solutions $A,D$ (red circles) 
and $B$ (blue diamonds) of Fig.~\ref{fig:per}(b).}
\label{fig:spec}
\end{figure}

\begin{figure}[tb]
\centering
\includegraphics[width=\columnwidth]{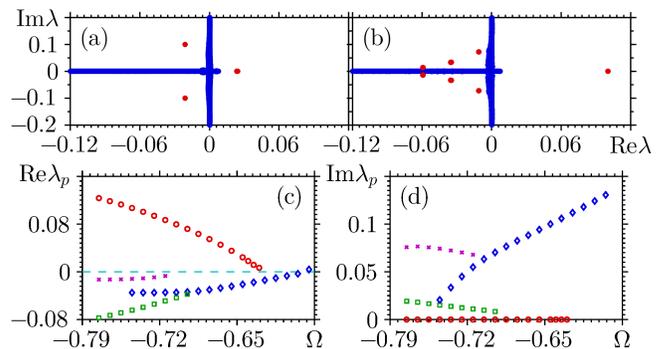}
\caption{(Color online) All eigenvalues [(a): $\W=-0.645$, (b): $\W=-0.735$, here $M=4096$, $N=64$] 
and point spectra in dependence on $\W$ (c,d) for the asymmetric
branch $C$.}
\label{fig:dspec}
\end{figure}

Below we present stability analysis for $\alpha=1.457$, for branches $A,B,C,D$ (see Fig.~\ref{fig:per}(b)). 
Four characteristic types of spectra are shown
in Fig.~\ref{fig:spec}(a-d). Only case (c) with point spectrum $\lambda_{p}$ having negative real part corresponds to a stable
chimera pattern, while all other patterns are unstable (oscillatory instability for cases (a,b) and monotonous instability 
for case (d)). Dependence of the point spectrum $\lambda_{p}$ on parameter $\W$ for $\alpha=1.457$, branches $A,B,D$, is 
shown in Fig.~\ref{fig:spec}(e,f). One can see that in the region  $-0.68\lesssim \W \lesssim-0.64$ there are four points
of $\lambda_{p}$, for other values of $\W$ there is only one pair of eigenvalues 
(or one real eigenvalue for branch $D$). This property may be attributed to the fact, that close to 
the homoclinic orbit $\W\approx\W_{\ast}$ the length of the patterns is large so here two discrete modes
are possible. The only stable chimera state is of type $A$ (we refer here to Figs.~\ref{fig:ill} and \ref{fig:per}(b))
with  $-0.91\lesssim \W \lesssim-0.69$.  
On the contrary, chimera states with two SRs (type $B$) are unstable.  Most difficult was analysis of two-SRs solutions of type~$C$~(Fig.~\ref{fig:dspec}), here the unstable branch of point spectrum $\lambda_{p}$ is real, and there is up to three stable complex pairs. In some cases only very fine discretization
with $M=6144$ allowed us to reveal unstable point eigenvalues~$\lambda_{p}$; we attribute this to a complex profile of this solution, requiring high resolution of perturbations.

\begin{figure}[tb]
\centering
\includegraphics[width=\columnwidth]{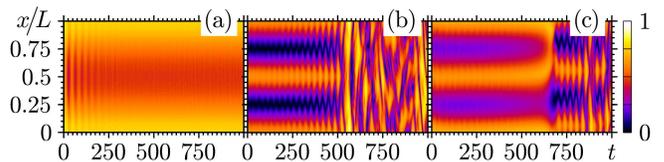}
\caption{\looseness=-1 (Color online) Direct numerical simulations of stable and unstable chimeras for $\alpha\!=\!1.457$. 
(a): Chimera of type $A$ for
$\W=-0.71377$, $L=6.03$;
(b): chimera of type $B$ for $\W=-0.575$, $L=12.06$; (c): chimera of type C, $\W=-0.6$, $L\!=\!12.06$. 
The number of oscillators was $K\!=\!700$ per length unit.}
\label{fig:simul}
\end{figure}

Stability properties have been confirmed
by direct numerical simulations of the ensemble governed by Eq.~\eqref{eq:kbmodel},{\,}\eqref{eq:kernel}, 
see Fig.~\ref{fig:simul} for space-time plots of field $|H(k,t)|=|\sum_{j}G(|k-j|/KL)\exp[i\phi_j]|$. One can initialize
all the chimera patterns found above; in the unstable regions these patterns are destroyed, while stable chimera persists.
Remarkably, for weakly unstable two-SRs chimeras for $\W\approx -0.58$, where the real part of the point eigenvalue $\lambda_{p}$
has a minimum (see Fig.~\ref{fig:spec}(e)), the life time of prepared chimera is relatively large.

Summarizing, in this Letter we reformulated the problem of chimera patterns in 1D medium
of coupled oscillators
as a system of PDEs. This allowed  to find uniformly rotating chimera states
as solutions of an ODE. Although a large variety of patterns with large spatial
periods can be found, we restricted our attention in this Letter to the simplest ones,  with at most
two synchronous domains.
Remarkably, these profiles can be explicitly described in the limit of neutral coupling between oscillators;
for coupling close to neutral one, a perturbation analysis yields approximate solutions. Exploring stability of the 
found solutions appeared to be a nontrivial numerical problem. We suggested an approach to characterize
the essential and the point parts of the spectrum via finite discretizations. It appears that only chimeras
of the type originally studied by Kuramoto and Battogtokh are stable, while other found patterns are 
linearly unstable.

The approach suggested could be extended in several directions. First, one can study general bifurcations
of chimera patterns. The difficulty here is that many tools for bifurcation analysis require sufficient smoothness
of the equations, what is not the case for chimera solutions. Stability analysis performed in this letter have been
restricted to perturbations with the same spatial period as the chimera itself, i.e. it describes stability for a medium
on a circle. Other unstable modes, e.g. of modulational instability type, 
could appear if one formulates the stability problem for an infinite medium. Finally, the formulated PDEs
have been simplified using the separation of time scales; it would be interesting to study stability of chimeras in the full
Eqs.~\eqref{eq:oa1},{\,}\eqref{eq:difeq1} with $\tau\neq 0$.\par
\acknowledgments
We acknowledge discussions with O.~Omelchenko, M.~Wolfrum, and Yu.~Maistrenko.
L.\,S. was supported by ITN COSMOS (funded by 
the European Union’s Horizon 2020 research and innovation
programme under the Marie Sklodowska-Curie grant agreement No 642563).
Numerical part of this work was  
supported by the Russian Science Foundation 
(Project No. 14-12-00811).\newpage
\bibliography{nld-old,nld-current,%
pap-ab,pap-ce,pap-fg,pap-hj,pap-kl,%
pap-mn,pap-oq,pap-rs,pap-tz,%
pik,books,n-stand,from_all_databases,%
cpkbm_note}

\begin{thebibliography}{18}
\bibitem [{bib({\natexlab{a}})}]{bib:cpkbm_note}%
  \BibitemOpen
  \href@noop {} {} \bibinfo {note} {{H}ere our definition of
  the frequency is the same as in the {KB} paper~[1]. This frequency will be
  negative, if $\alpha\lesssim \pi\bigl/2\bigr.$.}\BibitemShut {Stop}%
\end{thebibliography}
\end{document}